\documentclass[12pt]{extarticle}
\pdfoutput=1 
\usepackage{slashed}
\usepackage{latexsym}
\usepackage{amssymb}
\usepackage{amsmath}
\usepackage{diagbox} 
\usepackage{autobreak}
\usepackage{graphicx}
\usepackage{arydshln}
\usepackage{adjustbox}
\usepackage{easybmat}
\usepackage{bbm}
\usepackage{cite}
\usepackage{slashed}
\usepackage{bm}
\usepackage{adjustbox}
\usepackage{booktabs}
\usepackage{rotating}
\usepackage{mathtools}
\usepackage{lscape}
\usepackage{hyperref}
\usepackage{bm}
\usepackage{color}
\usepackage{fancybox,framed}
\usepackage{lscape}
\usepackage{multirow}
\usepackage{fancyhdr}
\usepackage{enumitem}
\usepackage{calrsfs}
\usepackage{tablefootnote}
\usepackage{cancel}
\usepackage[normalem]{ulem}
\usepackage[dvipsnames, svgnames, x11names]{xcolor}

\usepackage[most]{tcolorbox} 

\definecolor{block-gray}{gray}{0.95}

\definecolor{shadecolor}{rgb}{0.01,0.199,0.1}

\newtcolorbox{codeSyntax}{
    enhanced,
    frame hidden,
    colback=block-gray,
    boxrule=0pt,
    borderline west={2pt}{0pt}{gray!80!black}
}

\allowdisplaybreaks
\begin{document}
\thispagestyle{empty}
\begin{flushright}
{ANL--204508}
\end{flushright}
\vspace{0.8cm}

\begin{center}
{\Large\sc Renormalization of the SMEFT to dimension\\[0.5cm] eight: Fermionic interactions II}
\vspace{0.8cm}

\textbf{
Supratim Das Bakshi$^1$, Mikael Chala$^2$, and Zhe Ren$^2$
}\\
\vspace{1.cm}
{\em {$^1$ HEP Division, Argonne National Laboratory, Argonne, Illinois 60439, USA}}\\
\vspace{0.3cm}
{\em {$^2$ Departamento de F\'isica Te\'orica y del Cosmos,
Universidad de Granada, Campus de Fuentenueva, E--18071 Granada, Spain}}
\vspace{0.5cm}
\end{center}
\begin{abstract} 
We compute the one-loop mixing of bosonic and two-fermion interactions into two-fermion operators in the dimension-eight Standard Model Effective Field Theory (SMEFT). Together with the results in Refs.~\cite{Chala:2021pll,DasBakshi:2022mwk,Bakshi:2024wzz,Wu:2025qto}, this leaves only the mixing of four-fermion operators into two-fermion ones to complete the renormalization of the bosonic and two-fermion operators at this order.
\end{abstract}

\newpage

\section{Introduction}
The Standard Model (SM) extended with effective operators, commonly referred to as the SMEFT~\cite{Buchmuller:1985jz,Grzadkowski:2010es,Brivio:2017vri,Isidori:2023pyp,Aebischer:2025qhh}, provides a model-independent description of the low-energy limit of essentially any weakly-coupled scenario involving new physics at the TeV scale and above. As a result, the SMEFT is currently subject to intense experimental scrutiny using data from the Large Hadron Collider and low-energy facilities; see e.g. Refs.~\cite{Ellis:2018gqa,Straub:2018kue,Aoude:2020dwv,Boughezal:2020uwq,Dawson:2020oco,Anisha:2020ggj,Anisha:2021hgc,Alasfar:2022zyr,deBlas:2022hdk,Allwicher:2022gkm,Kassabov:2023hbm,Bartocci:2023nvp,Boughezal:2022pmb,Celada:2024mcf,Hammou:2024xuj,Bartocci:2024fmm,terHoeve:2025gey,deBlas:2025xhe,deBlas:2025hbr} for recent global fits.

This combined effort necessitates describing the theory across widely separated energy scales, which in turn requires knowledge of the renormalization group equations (RGEs) of the SMEFT. For lepton- and baryon-number–conserving interactions, the one-loop RGEs at dimension six---namely at $\mathcal{O}(v^2/\Lambda^2)$ in the expansion in powers of the electroweak scale $v\sim246$ GeV over the new physics cutoff $\Lambda$---were first computed in Refs.~\cite{Jenkins:2013zja,Jenkins:2013wua,Alonso:2013hga}. Partial two-loop results at this order in the effective field theory (EFT) have been subsequently obtained in Refs.~\cite{Panico:2018hal,Bern:2019wie,Bern:2020ikv,EliasMiro:2021jgu,Jenkins:2023bls,DiNoi:2024ajj,Banik:2025wpi,DiNoi:2025arz,DiNoi:2025tka,Duhr:2025zqw,Duhr:2025yor,Chala:2025crd,Haisch:2025lvd,Haisch:2025vqj,Duhr:2026btw} and culminated in Refs.~\cite{Born:2024mgz,Born:2026xkr}. We note that two-loop anomalous dimensions involving fermions can depend on the treatment of $\gamma_5$ in dimensional regularization. For example, Refs.~\cite{DiNoi:2025arz,Born:2026xkr} employ a 
naive dimensional regularization (NDR)/anticommuting prescription, while the relation between NDR and the BMHV scheme has been studied in Refs.~\cite{DiNoi:2025uan,Fuentes-Martin:2025meq}. This issue does not arise for the one-loop calculation performed in this work. There has also been remarkable progress in the renormalization of the low-energy EFT, or LEFT~\cite{Jenkins:2017jig}, both at one~\cite{Jenkins:2017dyc} and two-loops~\cite{Naterop:2024cfx,Naterop:2025lzc,Naterop:2025cwg}; as well as in the renormalization of more general EFTs~\cite{Fonseca:2025zjb,Misiak:2025xzq,Aebischer:2025zxg,Aebischer:2025ddl,Fonseca:2025cls,Guedes:2025sax}.

Likewise, given the potential relevance of dimension-eight interactions~\cite{Degrande:2020evl, Dawson:2021xei,  Dawson:2022cmu, Boughezal:2022nof, Heinrich:2022huh, Degrande:2023iob, Boughezal:2023nhe, Corbett:2023qtg, Ellis:2023zim, Celada:2024mcf, Assi:2024zap, deBlas:2025xhe}, the computation of the one-loop RGEs at $\mathcal{O}(v^4/\Lambda^4)$ has long been underway. Key results include the running of bosonic operators induced by pairs of dimension-six operators~\cite{Chala:2021pll} (see also Ref.~\cite{DasBakshi:2023htx}) and by dimension-eight operators themselves~\cite{DasBakshi:2022mwk}, the running of two-fermion operators induced by pairs of dimension-six operators~\cite{Bakshi:2024wzz}, and the running of four-fermion operators without Higgs fields or field strengths, $\psi^4 D^2$~\cite{Wu:2025qto}. (See also Ref.~\cite{AccettulliHuber:2021uoa} for pioneering work and Refs.~\cite{Helset:2022pde,Assi:2023zid,Ardu:2022pzk,Asteriadis:2022ras,Boughezal:2024zqa,Liao:2024xel,Henriksson:2025vyi,Chala:2025crd} for additional partial results, many of which have stimulated further research into structural aspects of EFTs, e.g. on the behavior of positivity bounds under running~\cite{Chala:2021wpj,Chala:2023jyx,Chala:2023xjy,Ye:2025zhs,Liao:2025npz}.)

In this work, we aim to advance this long-standing program by determining the renormalization of two-fermion operators induced by dimension-eight terms other than four-fermion ones. Since we perform the calculation off shell, this part of the analysis involves the largest set of redundant operators, including structures such as $\psi^2 X D^3$ and $\psi^2 \phi D^4$, which contribute to operators with many legs like $\psi^2 \phi^5$ through intricate and lengthy on-shell relations~\footnote{Ref.~\cite{Assi:2025fsm} contains some overlapping results, restricted to loops involving six- and seven-leg operators, which at one-loop cannot mix into operators with four or fewer legs. This considerably reduces the scope of the calculation compared to the present work.}.
We overcome this difficulty using a combination of \texttt{mosca}~\cite{LopezMiras:2025gar} and an ongoing update of \texttt{ABC4EFT}~\cite{Li:2020gnx,Li:2022tec,Li:2023cwy}. Both these relations and the final RGEs are available at \href{https://github.com/SMEFT-Dimension8-RGEs}{github.com/SMEFT-Dimension8-RGEs}.

This article is organized as follows. In section~\ref{sec:conventions}, we introduce our conventions. In section~\ref{sec:renormalization} we present the details of the calculation and summarize the main general results, including the off-shell and on-shell mixing patterns. We conclude in section~\ref{sec:discussion}, where we compare our findings with previous results in the literature and highlight the most significant conclusions.

\section{Conventions}\label{sec:conventions}
Following our previous work~\cite{Chala:2021pll,DasBakshi:2022mwk,Bakshi:2024wzz}, we use the following form of the SM Lagrangian:
\begin{align}\nonumber
 \mathcal{L}_\text{SM} = & -\frac{1}{4}G_{\mu\nu}^{A}G^{A\,\mu\nu} -\frac{1}{4}W_{\mu\nu}^{I}W^{I\,\mu\nu} -\frac{1}{4}B_{\mu\nu}B^{\mu\nu}\\\nonumber
 &
+\overline{q} i \slashed{D}q
+\overline{l} i \slashed{D}l
+\overline{u} i \slashed{D}u
+\overline{d} i \slashed{D}d
+\overline{e} i \slashed{D}e
\\
& +\left(D_{\mu}\phi\right)^{\dagger}\left(D^{\mu}\phi\right)
+\mu^{2}|\phi|^{2}-\lambda|\phi|^{4}
-\left(
\overline{q} \widetilde{\phi} y^{u} u 
+\overline{q}\phi y^{d} d
+\overline{l}\phi y^{e} e
+\text{h.c.}\right)~,
\end{align}
where $B$, $W$, and $G$ are the electroweak gauge bosons and the gluon, respectively; $l$, $q$ and $e$, $u$, $d$, represent the left-handed leptons and quarks and the right-handed counterparts, respectively; and $\phi$ stands for the Higgs doublet. Likewise, we use the minus-sign convention for the covariant derivative, with $g_1$, $g_2$ and $g_s$ denoting the $U(1)_Y$, $SU(2)_L$ and $SU(3)_c$ gauge couplings, respectively.

The lepton- and baryon-number-conserving SMEFT Lagrangian to dimension-eight reads:
\begin{equation}
    \mathcal{L}_\text{SMEFT} = \mathcal{L}_\text{SM} + \frac{1}{\Lambda^2}\sum_i c_i^{(6)}\mathcal{O}_i^{(6)} + \frac{1}{\Lambda^4}\sum_j c_{j}^{(8)}\mathcal{O}_j^{(8)}\,,
\end{equation}
where $\Lambda$ represents the new physics cutoff. We use the Warsaw basis of operators~\cite{Grzadkowski:2010es} at dimension-six and a slightly modified version of Murphy's basis~\cite{Murphy:2020rsh} at dimension-eight\footnote{Compared to Ref.~\cite{Murphy:2020rsh}, we use a different set of $\psi^2X^2D$ operators for simplicity, and we include some missing factors of the imaginary unit in $\psi^2X\phi^2D$ terms, to make them Hermitian. Furthermore, we change $\widetilde{\phi}^{\dagger} \overleftrightarrow{D}^\mu \phi$ to $\widetilde{\phi}^{\dagger} D^\mu \phi$ in all $udX\phi^2D$ operators, since the original $\mathcal{O}^{(1)}_{udWH^2}$ and $\mathcal{O}^{(2)}_{udWH^2}$ vanish. 
See \href{https://github.com/SMEFT-Dimension8-RGEs}{github.com/SMEFT-Dimension8-RGEs} for more details.}.

The RGEs that govern the dependence of Wilson coefficients (WCs) on the renormalization scale $\tilde{\mu}$ read:
\begin{equation}
    \dot{c}_i \equiv 16 \pi^2\tilde{\mu} \frac{d c_i^{(8)}}{d\tilde{\mu}} = \gamma_{ij}c_{j}^{(8)} + \gamma^\prime_{ijk}c_j^{(6)} c_k^{(6)}\,.
\end{equation}
For bosonic operators~\cite{Chala:2021pll,DasBakshi:2022mwk}, as well as for the four-fermion operators of class $\psi^4 D^2$~\cite{Wu:2025qto}, both $\gamma$ and $\gamma^\prime$ are known. For two-fermion ones, only $\gamma^\prime$ is known~\cite{Bakshi:2024wzz}. Here, we finally compute the remaining $\gamma$. \footnote{We leave the contributions from four-fermion dimension-eight operators, as well as the effects of two-fermion dimension-eight operators on the running of lower-dimensional operators, for future work.}

\section{Renormalization}\label{sec:renormalization}
As in our previous works~\cite{Chala:2021pll,DasBakshi:2022mwk,Bakshi:2024wzz}, and in contrast with Refs.\cite{AccettulliHuber:2021uoa,Wu:2025qto}, we do off-shell renormalization. We work in dimensional regularization with space-time dimension $D=4-2\epsilon$, and project the off-shell local divergences of one-particle-irreducible (1PI)
Feynman diagrams computed with \texttt{FeynRules}~\cite{Alloul:2013bka}, \texttt{FeynArts}~\cite{Hahn:2000kx}, \texttt{FormCalc}~\cite{Hahn:1998yk}, \texttt{FeynCalc}~\cite{Shtabovenko:2025lxq} and \texttt{Package-X}~\cite{Patel:2015tea} onto a Green's basis of SMEFT operators. We employ NDR for $\gamma_5$. In the present one-loop case, the calculation is well-defined and non-ambiguous.
We use the Green's basis of Ref.~\cite{Bakshi:2024wzz} extended with $\psi^2 D^5$, $\psi^2\phi D^4$, $\psi^2 X D^3$, $\psi^2 X \phi D^2$, $\psi^2 X^2 D$ and $\psi^2 X^2 \phi$. This is in turn a combination of the operators in Ref.~\cite{Chala:2021cgt} and those in Ref.~\cite{Ren:2022tvi} but in explicit Hermitian form; it can also be found at \href{https://github.com/SMEFT-Dimension8-RGEs}{github.com/SMEFT-Dimension8-RGEs}. We reduce redundant operators via equations of motion using automated tools, including \texttt{mosca}~\cite{LopezMiras:2025gar} and an updated version of \texttt{ABC4EFT}~\cite{Li:2022tec}, which will be made public in the near future.

In order to better illustrate our approach, we detail here the computation of the mixing of $\phi^6 D^2$ operators into $le \phi^5$. First, we notice that the on-shell equation for $le\phi^5$ (neglecting redundant operators which are not renormalized by $\phi^6 D^2$, see Tab.~\ref{tab:ADMred}) reads:
\begin{align}\label{eq:egonshellrelation}
    (c_{le\phi^5,mn})^{\text{on-shell}} &= (c_{le\phi^5,mn})^{\text{off-shell}} - r^{(3)}_{\phi^6D^2} y^e_{mn} + i r^{(4)}_{\phi^6D^2} y^e_{mn} + \frac{1}{4} g_1 r^{(3)}_{B\phi^4D^2} y^e_{mn} \nonumber \\
    & + \frac{1}{4} g_2 r^{(6)}_{W\phi^4D^2} y^e_{mn} + \frac{1}{8} g_2 r^{(7)}_{W\phi^4D^2} y^e_{mn} - \frac{1}{2} i r^{(2)}_{e^2\phi^4D,p n} y^e_{m p} + r^{(3)}_{e^2\phi^4D,p n} y^e_{m p} \nonumber \\
    & + \frac{i}{2} r^{(5)}_{l^2\phi^4D,m p} y^e_{p n} + r^{(6)}_{l^2\phi^4D,m p} y^e_{p n} + r^{(7)}_{l^2\phi^4D,m p} y^e_{p n} - i r^{(8)}_{l^2\phi^4D,m p} y^e_{p n}\,.
\end{align}
Although there are 1PI diagrams involving $\phi^6D^2$ vertices contributing to $le\phi^5$, the contribution turns out to vanish, so we have
\begin{align}(\dot{c}_{le\phi^5,mn})^{\text{off-shell}} = 0.
\end{align}

The only redundant operators in this equation that are (off-shell) renormalized by $\phi^6 D^2$ interactions are: 
\begin{align}\label{eq:egRGEoffshellf}
	\dot{r}^{(3)}_{\phi^6D^2} &= - \frac{3}{2} g_2^2 c^{(1)}_{\phi^6D^2}
	+ 3 g_2^2 c^{(2)}_{\phi^6D^2} + 2 \lambda c^{(1)}_{\phi^6D^2} - 4 \lambda c^{(2)}_{\phi^6D^2}\,,\\
	\dot{r}^{(3)}_{B\phi^4D^2} &= -\frac{1}{3} g_1 c^{(1)}_{\phi^6D^2} -\frac{2}{3} g_1 c^{(2)}_{\phi^6D^2}\,, \\
	\dot{r}^{(6)}_{W\phi^4D^2} &= -\frac{1}{3} g_2 c^{(1)}_{\phi^6D^2} +\frac{2}{3} g_2 c^{(2)}_{\phi^6D^2}\,, \\
	\dot{r}^{(7)}_{W\phi^4D^2} &= -\frac{4}{3} g_2 c^{(2)}_{\phi^6D^2}, \\
	\dot{r}^{(3)}_{e^2\phi^4D,mn} &= y^{e}_{pn} y^{e*}_{pm} c^{(1)}_{\phi^6D^2}\,, \\
	\dot{r}^{(6)}_{l^2\phi^4D,mn} &= \frac{1}{2}y^{e}_{mp} y^{e*}_{np} c^{(1)}_{\phi^6D^2}\,, \\
	\dot{r}^{(7)}_{l^2\phi^4D,mn} &=  \frac{1}{2}y^{e}_{mp} y^{e*}_{np} c^{(2)}_{\phi^6D^2}\,.
	\label{eq:egRGEoffshelll}
\end{align}

\begin{figure}[t]
    \centering
    \includegraphics[width=0.2\linewidth]{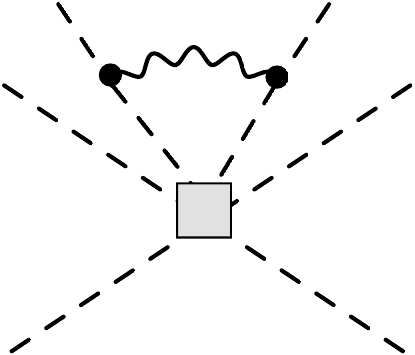}\hspace{1.5cm}
    \includegraphics[width=0.2\linewidth]{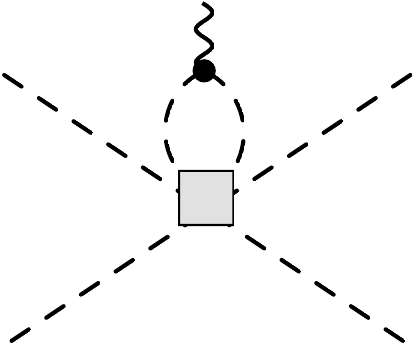}\hspace{1.5cm}
    \includegraphics[width=0.2\linewidth]{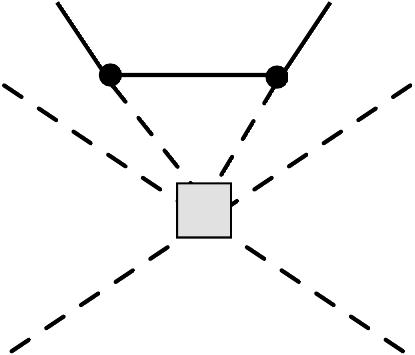}
    \caption{\it Example Feynman diagrams for the mixing of $\phi^6 D^2$ into $\phi^6 D^2$ (left), $\phi^4 X D^2$ (center) and $\psi^2\phi^4 D$ (right). The gray square denotes the effective interaction, while black dots represent SM couplings.}
    \label{fig:diagrams}
\end{figure}
Representative diagrams are shown in Fig.~\ref{fig:diagrams}.

Now, combining \eqref{eq:egonshellrelation} and \eqref{eq:egRGEoffshellf}--\eqref{eq:egRGEoffshelll}, we have the following total result:
\begin{align}\label{eq:comparison}
	(\dot{c}_{le\phi^5,mn})^{\text{on-shell}} &= \left(- \frac{1}{12} g_1^2 + \frac{17}{12} g_2^2 - 2 \lambda \right) y^e_{mn} c^{(1)}_{\phi^6D^2}  + \left(-\frac{1}{6} g_1^2 - 3 g_2^2 + 4 \lambda \right) y^e_{mn}c^{(2)}_{\phi^6D^2} \nonumber\\
	& + \frac{3}{2} y^e_{m r} y^e_{p n} y^{e*}_{p r} c^{(1)}_{\phi^6D^2}  + \frac{1}{2} y^e_{m r} y^e_{p n} y^{e*}_{p r} c^{(2)}_{\phi^6D^2} \,,
\end{align}
in perfect agreement with Ref.~\cite{Assi:2025fsm}.

We summarize the structure of mixing in four tables. In Tab.~\ref{tab:ADMred} we show which classes of operators (in columns) mix into others (in rows), including redundant ones. In Tab.~\ref{tab:reduction1} and Tab.~\ref{tab:reduction2} we describe how redundant classes (in columns) reduce to physical ones (in rows) on-shell, within our specific choice of Green's and physical bases. Finally, Tab.~\ref{tab:ADMphy} is the on-shell version of Tab.~\ref{tab:ADMred}. The actual RGEs can be found in \href{https://github.com/SMEFT-Dimension8-RGEs}{github.com/SMEFT-Dimension8-RGEs}. We have performed some cross-checks using \texttt{matchmakereft}~\cite{Carmona:2021xtq}.

\begin{table}[htbp]
	\centering
    \renewcommand{\arraystretch}{1.4}
	\resizebox{1.0\textwidth}{!}{
		\begin{tabular}{|*{12}{c|}}
			\hline
			& $\phi^8$ & $\phi^6 D^2$ & $X^2 \phi^4$ & $X \phi^4 D^2$ & $\phi^4 D^4$ & $\psi^2 \phi^5$ & $\psi^2 \phi^4 D$ & $\psi^2 X \phi^3$ & $\psi^2 \phi^3 D^2$ & $\psi^2 X \phi^2 D$ & $\psi^2 \phi^2 D^3$ \\
            \hline
			$\phi^6 D^2$ &
            $\smallsetminus$ &
            \checkmark &
            \checkmark &
            \checkmark &
            \checkmark &
            \checkmark &
            \checkmark &
            $\smallsetminus$ &
            \checkmark &
            $\smallsetminus$ &
            \checkmark \\
            \hline
			$X\phi^4 D^2$ &
            $\smallsetminus$ &
            \checkmark &
            0 &
            \checkmark &
            \checkmark &
            $\smallsetminus$ &
            \checkmark &
            0 &
            \checkmark &
            \checkmark &
            \checkmark \\
             \hline
			$\phi^4 D^4$ &
            $\smallsetminus$ &
            $\smallsetminus$ &
            $\smallsetminus$ &
            \checkmark &
            \checkmark &
            $\smallsetminus$ &
            $\smallsetminus$ &
            $\smallsetminus$ &
            \checkmark &
            $\smallsetminus$ &
            \checkmark \\
            \hline
			$X^2\phi^2 D^2$ &
            $\smallsetminus$ &
            $\smallsetminus$ &
            $\smallsetminus$ &
            \checkmark &
            \checkmark &
            $\smallsetminus$ &
            $\smallsetminus$ &
            $\smallsetminus$ &
            $\smallsetminus$ &
            \checkmark &
            \checkmark \\
            \hline
			\textcolor{gray}{$X\phi^2 D^4$} &
            $\smallsetminus$ &
            $\smallsetminus$ &
            $\smallsetminus$ &
            $\smallsetminus$ &
            \checkmark &
            $\smallsetminus$ &
            $\smallsetminus$ &
            $\smallsetminus$ &
            $\smallsetminus$ &
            $\smallsetminus$ &
            \checkmark \\
			\hline
			$\psi^2\phi^5$ &
            $\smallsetminus$ &
            \checkmark &
            \checkmark &
            0 &
            0 &
            \checkmark &
            \checkmark &
            \checkmark &
            \checkmark &
            0 &
            \checkmark \\
			\hline
			$\psi^2\phi^4D$ &
            $\smallsetminus$ &
            \checkmark & \checkmark & \checkmark & 0 & \checkmark &
            \checkmark & \checkmark &
            \checkmark &
            \checkmark & \checkmark \\
			\hline
			$\psi^2 X \phi^3$ &
            $\smallsetminus$ &
            $\smallsetminus$ & \checkmark &
            \checkmark &
            0 &
            0 &
            \checkmark &
            \checkmark &
            \checkmark &
            \checkmark &
            \checkmark \\
			\hline
			$\psi^2\phi^3D^2$ &
            $\smallsetminus$ &
            $\smallsetminus$ &
            $\smallsetminus$ &
            \checkmark &
            \checkmark &
            $\smallsetminus$ &
            \checkmark &
            \checkmark &
            \checkmark &            
            \checkmark &
            \checkmark \\
			\hline
			$\psi^2 X \phi^2 D$ & $\smallsetminus$ &
            $\smallsetminus$ &
            $\smallsetminus$ &
            \checkmark &
            \checkmark &
            $\smallsetminus$ &
            \checkmark &
            \checkmark &
            \checkmark &
            \checkmark &
            \checkmark \\
			\hline
			$\psi^2\phi^2D^3$ &
            $\smallsetminus$ &
            $\smallsetminus$ &
            $\smallsetminus$ &
            $\smallsetminus$ &
            \checkmark &
            $\smallsetminus$ &
            $\smallsetminus$ &
            $\smallsetminus$ &
            \checkmark &
            \checkmark &
            \checkmark \\
			\hline
			$\psi^2 X^2 \phi$ & $\smallsetminus$ & $\smallsetminus$ & $\smallsetminus$ & $\smallsetminus$ & $\smallsetminus$ & $\smallsetminus$ & $\smallsetminus$ & 0 & \checkmark & \checkmark & \checkmark \\
			\hline
			$\psi^2 X \phi D^2$ & $\smallsetminus$ & $\smallsetminus$ & $\smallsetminus$ & $\smallsetminus$ & $\smallsetminus$ & $\smallsetminus$ & $\smallsetminus$ & $\smallsetminus$ & \checkmark & \checkmark & \checkmark \\
			\hline
			\textcolor{gray}{$\psi^2\phi D^4$} & $\smallsetminus$ & $\smallsetminus$ & $\smallsetminus$ & $\smallsetminus$ & $\smallsetminus$ & $\smallsetminus$ & $\smallsetminus$ & $\smallsetminus$ & $\smallsetminus$ & $\smallsetminus$ & \checkmark \\
			\hline
			$\psi^2 X^2 D$ & $\smallsetminus$ & $\smallsetminus$ & $\smallsetminus$ & $\smallsetminus$ & $\smallsetminus$ & $\smallsetminus$ & $\smallsetminus$ & $\smallsetminus$ & $\smallsetminus$ & \checkmark & \checkmark \\
			\hline
			\textcolor{gray}{$\psi^2 X D^3$} & $\smallsetminus$ & $\smallsetminus$ & $\smallsetminus$ & $\smallsetminus$ & $\smallsetminus$ & $\smallsetminus$ & $\smallsetminus$ & $\smallsetminus$ & $\smallsetminus$ & $\smallsetminus$ & \checkmark \\
			\hline
			\textcolor{gray}{$\psi^2D^5$} & $\smallsetminus$ & $\smallsetminus$ & $\smallsetminus$ & $\smallsetminus$ & $\smallsetminus$ & $\smallsetminus$ & $\smallsetminus$ & $\smallsetminus$ & $\smallsetminus$ & $\smallsetminus$ & $\smallsetminus$ \\
			\hline
		\end{tabular}
	}
	\caption{The structure of the anomalous dimension matrix in the Green's basis, describing the off-shell mixing of operators in columns (only those that appear at tree level in SMEFT UV completions) into those in rows. Classes of operators in \textcolor{gray}{gray} are entirely redundant. $\smallsetminus$ means no (non-bubble) diagram contributes at one-loop, $0$ means there are diagrams but their combination vanishes, \checkmark \ means the contribution is non-vanishing.   This table is in agreement with the non-renormalization results of Refs.~\cite{Craig:2019wmo,Murphy:2020rsh}, which are however presented in the helicity basis.}
    \label{tab:ADMred}
\end{table}

\begin{table}[htbp]
	\centering
    \renewcommand{\arraystretch}{1.1}
	\resizebox{1.0\textwidth}{!}{
		\begin{tabular}{|*{9}{c|}}
			\hline
			& $\psi^2 \phi^4 D$ & $\psi^2 \phi^3 D^2$ & $\psi^2 X \phi^2 D$ & $\psi^2 \phi^2 D^3$ & $\psi^2 X \phi D^2$ & $\psi^2 \phi D^4$ & $\psi^2 X^2 D$ & $\psi^2 X D^3$  \\
			\hline
			$\psi^2\phi^5$ & \checkmark  & \checkmark & & \checkmark & & \checkmark & & \\
			\hline
			$\psi^2\phi^4D$ & & \checkmark & \checkmark & \checkmark & \checkmark & \checkmark & & \checkmark \\
			\hline
			$\psi^2 X \phi^3$ &  & \checkmark & \checkmark & \checkmark & \checkmark & \checkmark & & \checkmark \\
			\hline
			$\psi^2\phi^3D^2$ &  & \checkmark &  & \checkmark & \checkmark & \checkmark &  & \checkmark \\
			\hline
			$\psi^2 X \phi^2D$ &  &  & \checkmark & \checkmark & \checkmark & \checkmark & \checkmark & \checkmark \\
			\hline
			$\psi^2\phi^2D^3$ &  &  &  & \checkmark &  & \checkmark &  & \checkmark \\
			\hline
			$\psi^2 X^2 \phi$ &  &  &  &  & \checkmark & \checkmark & \checkmark & \checkmark \\
			\hline
			$\psi^2 X \phi D^2$ &  &  &  &  & \checkmark & \checkmark &  & \checkmark \\
			\hline
			$\psi^2 X^2 D$ &  &  &  &  &  &  & \checkmark & \checkmark \\
			\hline
		\end{tabular}
	}
	\caption{Redundant two-fermion operators (in columns) contributing to physical ones (in rows) upon field redefinitions. Only redundant operators renormalized by the tree-level generated operators in the SMEFT are included.}
    \label{tab:reduction1}
\end{table}

\begin{table}[htbp]
	\centering
    \renewcommand{\arraystretch}{1.4}
    \resizebox{0.7\textwidth}{!}{
		\begin{tabular}{|*{6}{c|}}
			\hline
			& $\phi^6 D^2$ & $X \phi^4 D^2$ & $\phi^4 D^4$ & $X^2 \phi^2 D^2$ & $X \phi^2 D^4$\\
			\hline
			$\psi^2\phi^5$ & \checkmark & \checkmark & \checkmark & \checkmark & \checkmark \\
			\hline
			$\psi^2\phi^4D$ &  & \checkmark & & \checkmark & \checkmark \\
			\hline
			$\psi^2 X \phi^3$ &  & & & & \\
			\hline
			$\psi^2\phi^3D^2$ &  &  & \checkmark &  & \checkmark \\
			\hline
			$\psi^2 X \phi^2D$ &  &  &  & \checkmark & \checkmark \\
			\hline
			$\psi^2\phi^2D^3$ &  &  &  &  & \checkmark \\
			\hline
			$\psi^2 X^2 \phi$ &  &  &  & \checkmark & \\
			\hline
			$\psi^2 X \phi D^2$ &  &  &  &  & \\
			\hline
			$\psi^2 X^2 D$ &  &  &  &  &  \\
			\hline
		\end{tabular}}
	\caption{Same as Tab.~\ref{tab:reduction1} but for redundant bosonic operators.}
    \label{tab:reduction2}
\end{table}

\section{Discussion}\label{sec:discussion}
This calculation almost completes the one-loop renormalization of the bosonic and two-fermion operators of the SMEFT to dimension eight; the only missing piece is the mixing of four-fermion operators into two-fermion ones. The running of the four-fermion operators with Higgs fields or field strengths ($\psi^4\phi^2$, $\psi^4 X$, $\psi^4\phi D$) is also still missing. (Certain $\mathcal{O}(v^2/\Lambda^2)$ corrections are also missing in the dimension-six sector; e.g., the running of dimension-six two-fermion terms due to loops of dimension-eight two-fermion interactions proportional to the Higgs mass $\mu^2$.)

Some comments on cross-checks are in order:
\begin{enumerate}
    \item Our results fulfill amplitude-based non-renormalization theorems~\cite{Cheung:2015aba,Murphy:2020rsh}. For example, $\psi^2 \phi^5$ operators can not mix into $\psi^2 \phi^4D$ ones on-shell (see Tab.~\ref{tab:ADMphy}), even though diagrammatically allowed; see Tab.~\ref{tab:ADMred}. 
    \item Our results fulfill the constraints implied by positivity restrictions~\cite{Chala:2023jyx,Chala:2023xjy}. For example, considering $\phi^4D^4$ and $l^2\phi^2D^3$ renormalizing $e^2\phi^2D^3$, and $\phi^4D^4$ and $e^2\phi^2D^3$ renormalizing $l^2\phi^2D^3$, in the one-flavour limit, we have:
    \begin{align}
    \begin{aligned}
        -\dot{c}_{e^2\phi^2 D^3}^{(1)}-\dot{c}_{e^2\phi^2 D^3}^{(2)}=&-\frac{1}{3} |y^{e}|^2 c^{(1)}_{\phi^4}-\frac{1}{2} |y^{e}|^2 c^{(2)}_{\phi^4}-\frac{1}{6} |y^{e}|^2 c^{(3)}_{\phi^4} \\& + \frac{2}{3} |y^{e}|^2 c^{(1)}_{l^2\phi^2D^3} + \frac{2}{3} |y^{e}|^2 c^{(2)}_{l^2\phi^2D^3}\,, \\
        -\dot{c}_{l^2\phi^2 D^3}^{(1)}-\dot{c}_{l^2\phi^2 D^3}^{(2)} + \dot{c}_{l^2\phi^2 D^3}^{(3)} + \dot{c}_{l^2\phi^2 D^3}^{(4)}=&-\frac{1}{6} |y^{e}|^2 c^{(1)}_{\phi^4}-\frac{1}{6} |y^{e}|^2 c^{(2)}_{\phi^4} \\& + \frac{1}{3} |y^{e}|^2 c^{(1)}_{e^2\phi^2D^3} + \frac{1}{3} |y^{e}|^2 c^{(2)}_{e^2\phi^2D^3}\,, \\
        -\dot{c}_{l^2\phi^2 D^3}^{(1)}-\dot{c}_{l^2\phi^2 D^3}^{(2)} - \dot{c}_{l^2\phi^2 D^3}^{(3)} - \dot{c}_{l^2\phi^2 D^3}^{(4)}=&-\frac{1}{6} |y^{e}|^2 c^{(1)}_{\phi^4}-\frac{1}{3} |y^{e}|^2 c^{(2)}_{\phi^4}-\frac{1}{6} |y^{e}|^2 c^{(3)}_{\phi^4} \\& + \frac{1}{3} |y^{e}|^2 c^{(1)}_{e^2\phi^2D^3} + \frac{1}{3} |y^{e}|^2 c^{(2)}_{e^2\phi^2D^3}\,.
    \end{aligned}
    \end{align}
    Using the rotation matrices $P_O$ defined in the ancillary file in Ref.~\cite{Chala:2023xjy}~\footnote{Specifically: $P_{e^2\phi^2 D^3}=\begin{bmatrix}2 & 3 \\-3 & -3 \end{bmatrix}$, $P_{\phi^4}=\begin{bmatrix}-1 & 1 & 0 \\1 & 0 & 0 \\ 0 & -1 & 1 \\ \end{bmatrix}$, $P_{l^2\phi^2 D^3}=\begin{bmatrix}2 & 3 & 4 & 5 \\-\frac{5}{2} & -\frac{7}{2} & -4 & -5 \\ 4 & 5 & 6 & 7 \\ -\frac{7}{2} & -\frac{11}{2} & -6 & -7\end{bmatrix}$.} such that $\vec{c}_O = P_O\cdot\vec{\tilde{c}}_O$, we obtain the following equations:
    \begin{align}
    \begin{aligned}
        \dot{\tilde{c}}_{e^2\phi^2 D^3}^{(1)}&=-\frac{1}{6} |y^{e}|^2 \tilde{c}^{(1)}_{\phi^4}-\frac{1}{6} |y^{e}|^2 \tilde{c}^{(2)}_{\phi^4}-\frac{1}{6} |y^{e}|^2 \tilde{c}^{(3)}_{\phi^4} - \frac{1}{3} |y^{e}|^2 \tilde{c}^{(1)}_{l^2\phi^2D^3} - \frac{1}{3} |y^{e}|^2 \tilde{c}^{(2)}_{l^2\phi^2D^3}\,, \\
        \dot{\tilde{c}}_{l^2\phi^2 D^3}^{(1)}&=-\frac{1}{6} |y^{e}|^2 \tilde{c}^{(2)}_{\phi^4} - \frac{1}{3} |y^{e}|^2 \tilde{c}^{(1)}_{e^2\phi^2D^3}\,, \\
        \dot{\tilde{c}}_{l^2\phi^2 D^3}^{(2)}&=-\frac{1}{6} |y^{e}|^2 \tilde{c}^{(1)}_{\phi^4} -\frac{1}{6} |y^{e}|^2 \tilde{c}^{(3)}_{\phi^4} - \frac{1}{3} |y^{e}|^2 \tilde{c}^{(1)}_{e^2\phi^2D^3}\,.
    \end{aligned}
    \end{align}
    All anomalous dimensions in this basis are negative-definite, in agreement with the results in Ref.~\cite{Chala:2023xjy}, obtained solely from unitarity and analyticity of the S-matrix.
    \item
     Our results match those obtained in Ref.~\cite{Assi:2025fsm} in the overlapping cases (see e.g. Eq.~\ref{eq:comparison} above), with the exception of the mixing of $\psi^2 \phi^4 D$ into $\psi^2 \phi^5$ and into $\psi^2 \phi^4 D$ itself, where our results differ from theirs in some terms proportional to $g_2^2$.\footnote{We have independently reproduced this specific mixing using the (\href{https://gitlab.com/matchete/matchete/-/blob/master/Models/SMEFT-D8.m?ref_type=heads}{currently undocumented}) \texttt{UVDivergentAction} function of \texttt{matchete}~\cite{Fuentes-Martin:2022jrf}, finding agreement with our result.}. Note though that Ref.~\cite{Assi:2025fsm} uses a slightly different basis of operators: It is Murphy's basis~\cite{Murphy:2020rsh} with changes in $\psi^2\phi^4 D$ terms, such that ours ($c$) are related to theirs ($\tilde{c})$ by:
    \begin{align}
        c_{l^2\phi^4 D}^{(2)} = \frac12 (\tilde{c}_{l^2\phi^4 D}^{(2)}+\tilde{c}_{l^2\phi^4 D}^{(3)})\,,\quad c_{l^2\phi^4 D}^{(3)} = \tilde{c}_{l^2\phi^4 D}^{(4)}\,,\quad c_{l^2\phi^4 D}^{(4)} = \frac12 (\tilde{c}_{l^2\phi^4 D}^{(3)}-\tilde{c}_{l^2\phi^4 D}^{(2)})\,, 
    \end{align}
    while $c_{l^2\phi^4 D}^{(1)}$ remains the same. Analogous relations hold for the corresponding quark operators.
\end{enumerate}

\begin{table}[t]
	\centering
    \renewcommand{\arraystretch}{1.4}
	\resizebox{1.0\textwidth}{!}{
		\begin{tabular}{|*{12}{c|}}
			\hline
			& $\phi^8$ & $\phi^6 D^2$ & $X^2 \phi^4$ & $X \phi^4 D^2$ & $\phi^4 D^4$ & $\psi^2 \phi^5$ & $\psi^2 \phi^4 D$ & $\psi^2 X \phi^3$ & $\psi^2 \phi^3 D^2$ & $\psi^2 X \phi^2 D$ & $\psi^2 \phi^2 D^3$ \\
            \hline
			$\psi^2\phi^5$ &
            $\smallsetminus$ &
            \checkmark &
            \checkmark &
            \checkmark &
            \checkmark &
            \checkmark &
            \checkmark &
            \checkmark &
            \checkmark &
            \checkmark &
            \checkmark \\
			\hline
			$\psi^2\phi^4D$ & $\smallsetminus$ & \checkmark & 0 & \checkmark & \checkmark & 0 & \checkmark & 0 & \checkmark & \checkmark & \checkmark \\
			\hline
			$\psi^2 X \phi^3$ &
            $\smallsetminus$ &
            0 & \checkmark &
            \checkmark & \checkmark & 0 & 0  & \checkmark & \checkmark & \checkmark & \checkmark \\
			\hline
			$\psi^2\phi^3D^2$ & $\smallsetminus$ & $\smallsetminus$ & $\smallsetminus$ & \checkmark & \checkmark & $\smallsetminus$ & 0 & 0 & \checkmark & \checkmark & \checkmark \\
			\hline
			$\psi^2 X \phi^2 D$ & $\smallsetminus$ & $\smallsetminus$ & $\smallsetminus$ & \checkmark  & \checkmark  & $\smallsetminus$ & 0 & 0 & \checkmark & \checkmark & \checkmark \\
			\hline
			$\psi^2\phi^2D^3$ & $\smallsetminus$ & $\smallsetminus$ & $\smallsetminus$ & $\smallsetminus$ & \checkmark & $\smallsetminus$ & $\smallsetminus$ & $\smallsetminus$ & 0 & 0 & \checkmark \\
			\hline
			$\psi^2 X^2 \phi$ & $\smallsetminus$ & $\smallsetminus$ & $\smallsetminus$ & 0 & 0 & $\smallsetminus$ & $\smallsetminus$ & 0 & 0 & \checkmark & \checkmark \\
			\hline
			$\psi^2 X \phi D^2$ & $\smallsetminus$ & $\smallsetminus$ & $\smallsetminus$ & $\smallsetminus$ & 0 & $\smallsetminus$ & $\smallsetminus$ & $\smallsetminus$ & 0 & 0 & \checkmark \\
			\hline
			$\psi^2 X^2 D$ & $\smallsetminus$ & $\smallsetminus$ & $\smallsetminus$ & $\smallsetminus$ & $\smallsetminus$ & $\smallsetminus$ & $\smallsetminus$ & $\smallsetminus$ & $\smallsetminus$ & 0 & \checkmark \\
			\hline
		\end{tabular}
	}
	\caption{Same as Tab.~\ref{tab:ADMred} but in the physical basis.}
    \label{tab:ADMphy}
\end{table}

Given the length and technical complexity of the calculation, it will be valuable for the results presented here to be further scrutinized and independently reproduced.

\section*{Acknowledgments}
We are very thankful to Andreas Helset and the authors of Ref.~\cite{Assi:2025fsm} for their helpful assistance in comparing our results with those presented in that work. We also thank Radja Boughezal, Renato Fonseca, Xu Li, and Frank Petriello for useful discussions. MC is supported by the European Research Council under grant agreement n. 101230200. This work has received further funding from MICIU/AEI/10.13039/501100011033
(grants PID2022-139466NB-C21/C22 and PID2024-161668NB-100), from European Union NextGenerationEU/PRTR under grant CNS2022-136024 as well as from Junta de Andaluc\'ia (grants FQM 101 and P21-00199). SDB is supported by the U.S.~Department of Energy under contract DE-AC02-06CH11357. 

\bibliographystyle{style}
\bibliography{refs}

\end{document}